\documentclass{elsart}
\usepackage{graphicx}

\def\simgreat{{\th \rlap{\raise 0.5ex\hbox{$\scriptstyle  {>}$}}
    {\lower 0.3ex\hbox{$\scriptstyle  {\sim}$}} \th }}  
\def\th{\thinspace}

\begin{document}

\begin{frontmatter}

\title{Performance analysis of direct $N$-body algorithms 
for astrophysical simulations on distributed systems}

\author[AS]{Alessia Gualandris}, 
\author[AS]{Simon Portegies Zwart} and 
\author[CS]{Alfredo Tirado-Ramos}

\address[AS]{Astronomical Institute and Section Computational Science, 
  University of Amsterdam, the Netherlands.}
\address[CS]{Section Computational Science, University of Amsterdam, the Netherlands.}

\begin{abstract}
We discuss the performance of direct summation codes used in the
simulation of astrophysical stellar systems on highly distributed
architectures.  These codes compute the gravitational interaction
among stars in an exact way and have an $O(N^2)$ scaling with the
number of particles.  They can be applied to a variety of
astrophysical problems, like the evolution of star clusters, the
dynamics of black holes, the formation of planetary systems, and
cosmological simulations.  The simulation of realistic star clusters
with sufficiently high accuracy cannot be performed on a single
workstation but may be possible on parallel computers or grids.  We
have implemented two parallel schemes for a direct $N$-body code and
we study their performance on general purpose parallel computers and
large computational grids.  We present the results of timing analyzes
conducted on the different architectures and compare them with the
predictions from theoretical models.  We conclude that the simulation
of star clusters with up to a million particles will be possible on
large distributed computers in the next decade.  Simulating entire
galaxies however will in addition require new hybrid methods to
speedup the calculation.
\end{abstract}

\end{frontmatter}

\section{Introduction}
Numerical methods for solving the classical astrophysical $N$-body
problem have evolved in two main directions in recent years.  On the
one hand, approximated models like Fokker-Planck models
\cite{1980ApJ...242..765C}, \cite{1991ApJ...370...60M}, gaseous models
\cite{1991MNRAS.251..408L}, and Monte Carlo models
\cite{1975IAUS...69..133H}, \cite{1975IAUS...69....3S},
\cite{1998MNRAS.298.1239G} have been applied to the simulation of
globular clusters and galactic nuclei.  These models permit to follow
the global evolution of large systems along their lifetime but at the
expense of moderate accuracy and resolution. The basic approach is to
group particles according to their spatial distribution and use a
truncated multipole expansion to evaluate the force exerted by the
whole group instead of evaluating directly the contributions from the
single particles.  On the other hand, direct summation methods have
been developed to accurately model the dynamics and evolution of
collisional systems like dense star clusters. These codes compute all
the inter-particle forces and are therefore the most accurate. They
are also more general, as they can be used to simulate both low and
high density regions.  Their high accuracy is necessary when studying
physical phenomena like mass segregation, core collapse, dynamical
encounters, formation of binaries or higher order systems, ejection of
high velocity stars, and runaway collisions.  Direct methods have an
$O(N^2)$ scaling with the number of stars and are therefore limited to
smaller particle numbers compared to approximated methods.  For this
reason, so far they have only been applied to the simulation of
moderate size star clusters.  The simulation of globular clusters
containing one million stars is still a challenge from the
computational point of view, but it is an important astrophysical
problem.  It will provide insight in the complex dynamics of these
collisional systems, in the microscopic and macroscopic physical
phenomena, and it will help finding evidence of the controversial
presence of a central black hole.  The simulation of the evolution of
these systems under the combined effects of gravity, stellar evolution,
and hopefully hydrodynamics will allow to study the stellar population
and to compare the results with observations.

The need to simulate ever larger systems and to include a realistic
physical treatment of stars asks for a speedup in the most demanding
part of the calculation: the gravitational dynamics.  Significant
improvement in the performance of direct codes can be obtained either
by means of special purpose computers like GRAPE hardware
\cite{2003PASJ...55.1163M} or of general purpose distributed systems
\cite{775815}.  In this work, we focus on the performance of direct
$N$-body codes on distributed systems.  Two main classes of algorithms
can be used to parallelize direct summation $N$-body codes: replicated
data and distributed data algorithms.  In this work we present a
performance analysis of the two algorithms on different architectures:
a Beowulf cluster, three supercomputers, and two computational grids.
We provide theoretical predictions and actual measurements for the
execution time on different platforms, allowing the choice of the best
performing scheme for a given architecture.

\section{Numerical method}
In the direct method the gravitational force acting on a particle is
computed by summing up the contributions from all the other particles
according to Newton's law
\begin{equation}
\mathbf F_i = m_i \mathbf a_i = -G m_i \sum_{j=1, j\neq i}^{j=N}
\frac{m_j(\mathbf r_i-\mathbf r_j)}{|\mathbf r_i-\mathbf r_j|^3}.
\end{equation}
The number of force calculations per particle is $N(N-1)/2$ .  Given
the fact that the force acting on a particle usually varies smoothly
with time, the integration of the particle trajectory makes use of
force polynomial fitting.  In this work we implement the fourth-order
Hermite integrator with a predictor-corrector scheme and a
hierarchical time-step.

In the Hermite scheme \cite{1992PASJ...44..141M} high order
derivatives are explicitly computed to construct interpolation
polynomials of the force.  After the group of particles to be
integrated at time $t$ has been selected, the positions and velocities
of all particles are predicted at time $t$ ({\it predictor phase})
using the values of positions, velocities, accelerations, and first
derivative of accelerations (hereafter jerks) computed at the previous
step.  The prediction uses a third order Taylor expansion.  By means
of the predicted quantities, new values of the accelerations and jerks
at time $t$ are computed. This calculation is the most computationally
expensive of the whole scheme, having a $N^2$ scaling. The second and
third derivative of the accelerations are then calculated using the
Hermite interpolation based on the values of acceleration and
jerk. These correcting factors are added to the predicted positions
and velocities ({\it corrector phase}) at time $t$. The new time-step
of the particles is estimated according to the time-step prescription
in use and the time of the particles is updated.

The hierarchical time-step scheme is a modification of the individual
time-step scheme in which groups of particles are forced to share the
same time-step.  In the {\it individual time-step} scheme every
particle has its own time $t_i$ and its own time-step $\Delta t_i$,
with a step-size depending on the value of the acceleration $a_i$ and
its higher order derivatives $a_i^{(n)}$ according to
\begin{equation}
\Delta t_i = \sqrt{\eta \frac{|\mathbf a_i||\mathbf a_i^{(2)}|+
    |\mathbf a_i^{(1)}|^2}{|\mathbf a_i^{(1)}||\mathbf a_i^{(3)}|+|\mathbf a_i^{(2)}|^2}},
\end{equation}
where $\eta$=0.01 is a dimensionless accuracy parameter.  Only
particles requiring a short time-step are integrated with such a short
step-size, while other particles can be integrated with a longer one.
This reduces the total calculation cost by a factor $O(N^{1/3})$ with
respect to a shared time-step code, where all the particles share the
same time-step \cite{1988ApJS...68..833M}.  However, it is not
efficient to use the individual time-step scheme in its original form
on a parallel computer since only one particle is integrated at each
step, that is the particle with the smallest value of $t_i + \Delta
t_i$.  In order to fully exploit a parallel code, several particles
need to be updated at the same time, or, equivalently, several
particles need to share the same time-step.  In the {\it hierarchical}
or {\it block time-step} scheme \cite{MM86} the time-steps are
quantized to powers of two, so that a group of particles can be
advanced at the same time.  After the computation of $\Delta t_i$
according to the individual time-step prescription, the largest power
of two smaller than $\Delta t_i$ is actually assigned as a time-step
to the particle under consideration. The group of particles sharing
the same time-step are said to form a {\it block}.  The use of block
time-steps results in a better performance since it permits to advance
several particles simultaneously: the computation of the force exerted
upon the particles in a block and the integration of the trajectories
can be done in parallel by different processors.  Furthermore, the
positions and velocities in the predictor phase need to be calculated
only once for these particles.

\section{Parallel schemes for direct $N$-body codes}
The parallelization of a direct $N$-body code can proceed in different
ways depending on the desired intrinsic degree of parallelism and
communication to computation ratio.  We implemented two different
parallelization algorithms, the {\it copy algorithm} and the {\it ring
algorithm}, for an Hermite scheme with block time-steps using the
standard MPI library package.  If we denote with $N$ the total number
of particles in the system and with $p$ the number of available
processors, both algorithms have a theoretical computational
complexity $O(Np)$ for the communication and $O(N^2/p)$ for the
calculation (see $\S$ \ref{sec:parallel} for a derivation of more
detailed scaling relations).

\subsection{The copy algorithm}
\begin{figure}
\centering
 \includegraphics[width=7.5cm]{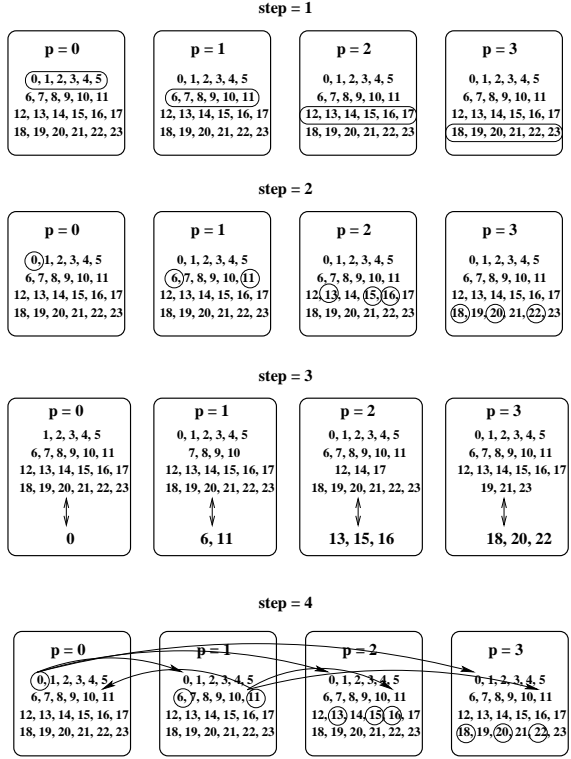}
 \caption{Schematic representation of the force calculation in the
    copy algorithm in the simple case of $p$=4 processors and $N$=24
    particles. Each processor has a local copy of the whole system but
    at step 1 is also assigned $n$=6 particles (selected particles).
    Among these, at step 2 each processor selects the local subgroup
    $s_i$ of the block of particles to be updated (circled particles).
    At step 3 each node calculates the total force exerted by all the
    $N$ particles (those on top of the arrow) on the selected
    particles (those below the arrow). At step 4 every processor
    broadcasts the new data to all the other processors. For clarity
    only the communication operations (indicated with arrows) of
    processor 0 and 1 are shown in the figure.}
 \label{fig:copyfig}
\end{figure}

The copy algorithm \cite{2002NewA....7..373M}, also called the
``replicated data algorithm'' \cite{775815}, is a parallelization
scheme which relies on the assumption that each processor has a local
copy of the whole system.  A schematic representation of the copy
algorithm is shown in Fig.\,\ref{fig:copyfig}.
At each integration step, the particles to be advanced are distributed
among the processors. The nodes proceed in parallel to the computation
of the forces exerted by all the $N$ particles on the local particles,
of the trajectories, and of the new time-steps.  At the end of the step
all the processors broadcast the new data to all the other processors
for a complete update.

This algorithm minimizes the amount of communication since it only
requires one collective communication at the end of each step.  The
main disadvantage is its limitation in memory because of the need to
store the data relative to all the particles on each node\footnote{
Our code requires approximately 124 bytes for each particle, which
amounts to about 130 Mbytes for a million particles system.  This
requirement is not prohibitive for present PCs, but becomes important
in the case of a parallel setup of GRAPE-6A, for which a maximum of
128 thousands particles can be stored.}.  This scheme may suffer from
load imbalance as the number of particles to update can be different
for the different nodes.  For high numbers of particles, however, a
random initial distribution is sufficient to ensure a relatively good
balance.

\subsection{The ring algorithm}
\begin{figure}
  \begin{center}
    \includegraphics[width=6cm]{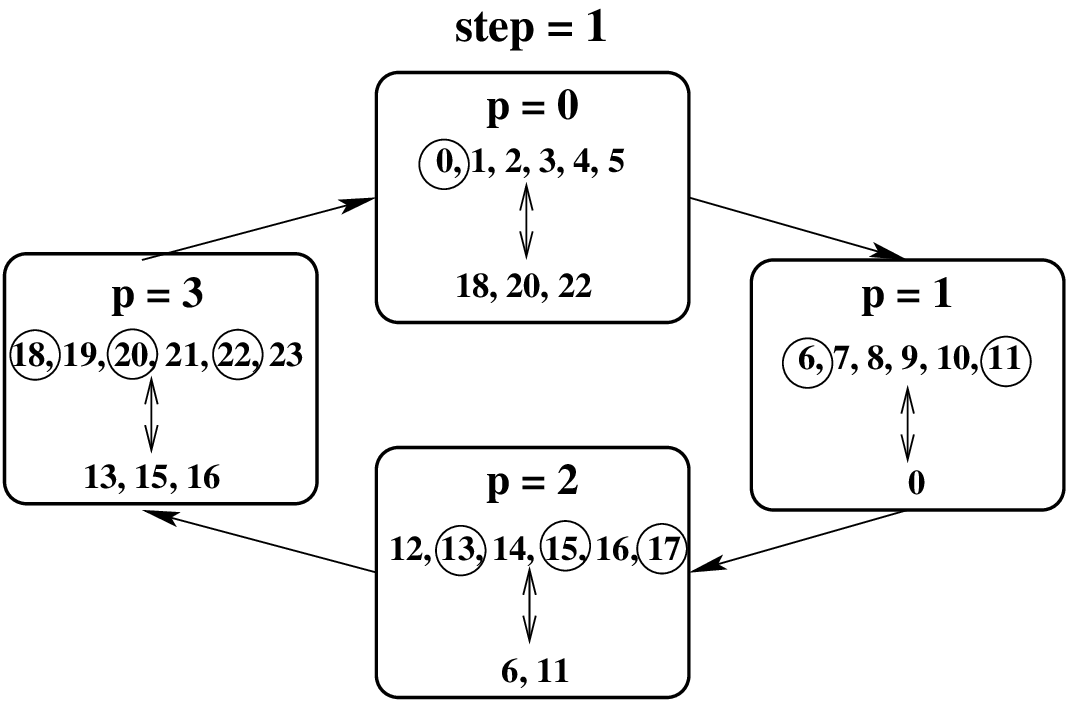}
    \includegraphics[width=6cm]{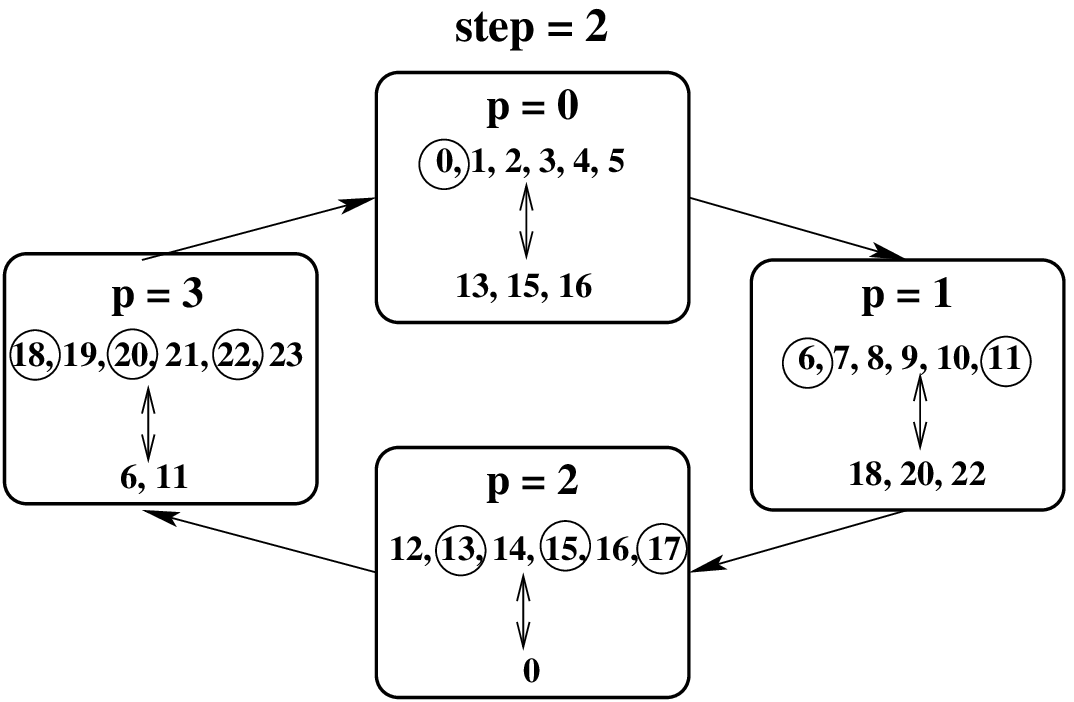}
    \includegraphics[width=6cm]{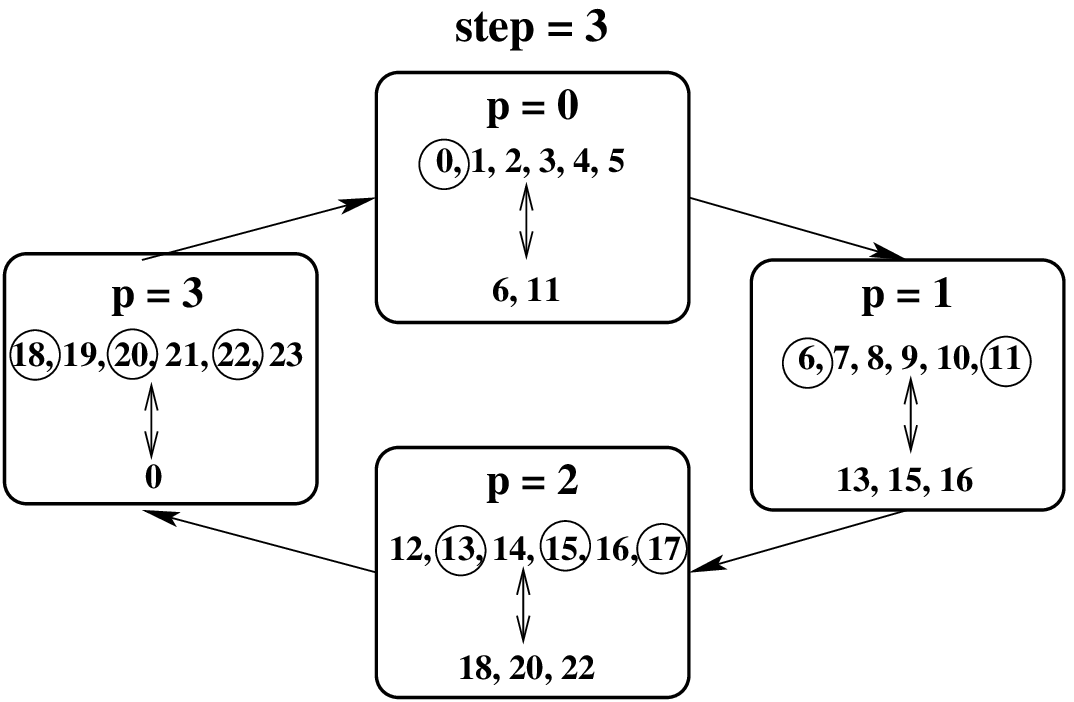}
    \includegraphics[width=6cm]{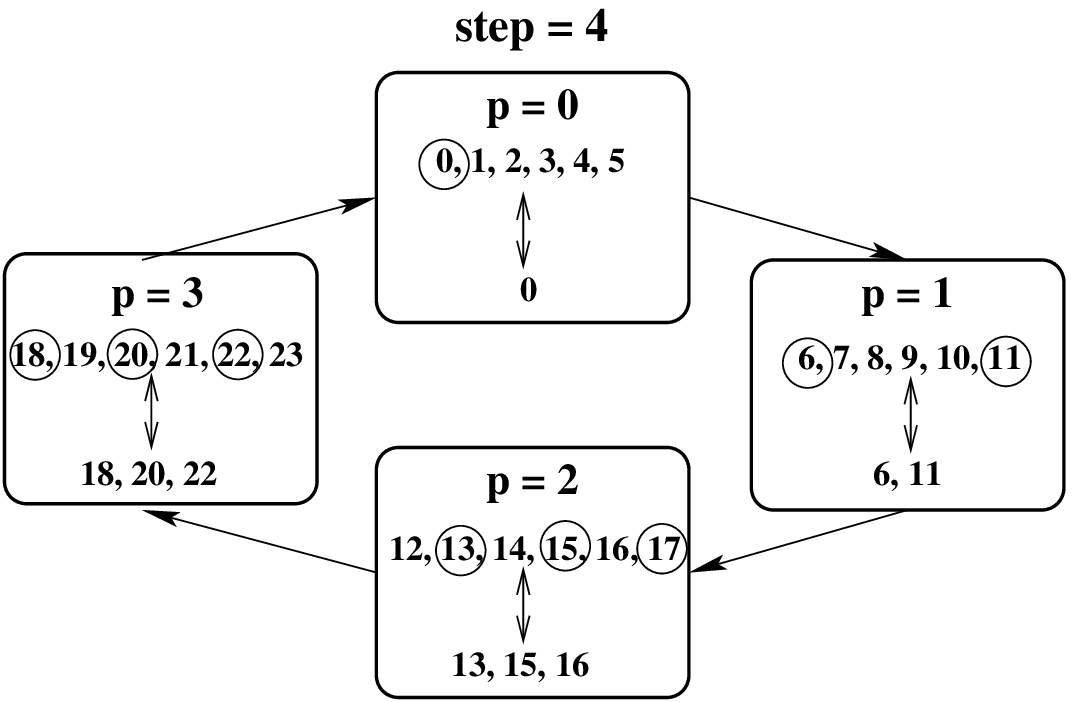}
    \caption{Schematic representation of the force calculation in the
      ring algorithm in the simple case of $p$=4 processors and $N$=24
      particles.  The processors are virtually connected in a
      ring-like structure topology.  At the beginning every processor
      has $n$=6 particles and a subset $s_i$ of them needs to be
      updated. Each node starts computing the partial forces exerted
      by its $n$ particles on those in the block.  At step 1 each
      processor sends the data relative to the $s_i$ particles
      (circled particles in the first row) to its right neighbor,
      receives particles (second row) from the left neighbor and
      computes the force on the received particles (those below the
      arrow).  The force computation is indicated with vertical
      arrows.  The same procedure is repeated in step 2 and 3 until in
      step 4 the total forces on the block of particles are computed
      and the $s_i$ particles are returned to their owners.}
    \label{fig:ringfig}
  \end{center}
\end{figure}

The ring or systolic algorithm makes use of a virtual ring topology of
the processors to circulate the particles which need to be advanced.
In this scheme, each processor is assigned a group of $n=N/p$
particles and only needs to store the data relative to those particles
throughout the whole integration. Therefore, the scheme has the
advantage of minimizing the memory requirements on each node.  A
schematic representation of the ring algorithm is shown in
Fig.\,\ref{fig:ringfig}.

Among all the particles that need to be updated, each processor
selects those belonging to its local group.  The initialization, the
predictor phase and the corrector phase are performed in parallel by
all the processors on their local subgroup of particles. When the
acceleration on the subgroup needs to be computed, each node first
calculates the partial forces exerted by all the $n$ particles and
then sends the data (including the partial acceleration) to the
processor which is defined as the right neighbor in the virtual
topology.  At the same time, the node receives data from its left
neighbor and starts incrementing the accelerations exerted by its
local $n$ particles on the received ones. After $p$ shifts, the
complete forces are calculated and the particles are returned to their
original processors for the computation of the trajectory and the
update of the time and time-step values.

The ring algorithm can be implemented using MPI non-blocking
communication routines \cite{775815}.  This technique is called {\it
latency hiding} and is most efficient in the case of small number of
particles or of very concentrated models, when the block sizes are
smaller and hence the load imbalance becomes more important.
Non-blocking MPI communication routines allow the separation between
the initiation and the completion of a communication by returning
immediately after the start of the communication.  In this way, some
computation can be performed at the same time as the sending and
receiving of data is ongoing.  The implementation of non-blocking
communication in our Hermite code exploits the property that the
computation of the force only requires the positions and velocities of
the particles.  The transfer of positions and velocities can then be
separated from the transfer of accelerations and jerks.  In each shift
of the systolic scheme, the communication is split in two branches and
overlaps with the computation of new forces and their derivatives. The
branch of accelerations and jerks follows one step behind that of
positions and velocities.  As a consequence of that, the transfer of
the forces starts only at the second shift and one final transfer is
necessary at the end of the last shift.

\section{Performance analysis of different parallel $N$-body schemes}
\label{sec:parallel}
The performance of a parallel code does not only depend on the
properties of the code itself, like the parallelization scheme and the
intrinsic degree of parallelism, but also on the properties of the
parallel computer used for the computation.  The main factors
determining the general performance are the calculation speed of each
node, the bandwidth of the inter-processor communication, and the
start-up time ({\it latency}).  A theoretical estimate of the total
time needed for one full force calculation loop must take into account
the properties of the parallel computer \cite{775815},
\cite{2002NewA....7..373M}.  Table\,\ref{tab:computer} shows the
hardware specifications of the different platforms used for our
performance runs: a Beowulf cluster, a SGI Origin supercomputer
(Teras), an experimental low latency cluster (DAS-AMS), and a
state-of-the-art computer cluster (Lisa).
\begin{table}[h]
  \caption{Specifications for the different distributed architectures.}
  \label{tab:computer}
  \centering
    \begin{tabular}{ccccc}
      \hline
      & Beowulf  & Teras    &  DAS-AMS  & Lisa\\
      \hline
      \hline
      OS & Linux & Irix & Linux & Linux \\
      Compiler & gcc-2.95.4 & gcc-3.0.4 & gcc-2.96 & gcc-3.3\\
      CPU & AMD Athlon & MIPS R14000 & Pentium-III & Intel Xeon\\
      CPU speed & 700 MHz &  500 MHz & 1 GHz & 3.4 GHz\\
      Network &  Fast Ethernet  & Gbit Ethernet  & Myrinet-2000 & Infiniband  \\
      $\tau_f$ [sec] & 5.5$\times10^{-7}$ & 3.5$\times10^{-7}$ & 4.5$\times10^{-7}$  & 1.2$\times10^{-7}$\\
      $\tau_l$ [sec] & 5.0$\times10^{-5}$ & 2.0$\times10^{-5}$ & 1.0$\times10^{-5}$  & 5.0$\times10^{-6}$\\
      $\tau_c$ [sec] & 8.0$\times10^{-4}$ & 1.5$\times10^{-4}$ & 1.5$\times10^{-4}$  & 5.0$\times10^{-5}$\\
      \hline
    \end{tabular}
\end{table}

The Beowulf cluster and Teras are hosted by the SARA computing and
networking services center in Amsterdam.  DAS-2 is a wide-area
distributed computer composed by clusters located at five different
universities in the Netherlands.  For the tests described in this
section we only use the nodes in Amsterdam (DAS-AMS) which are
interconnected by a fast and low latency network.  Lisa is the Dutch
national computer cluster hosted by SARA.  In the Table we report the
specifications, the time $\tau_f$ for the computation of one
inter-particle force, the time $\tau_l$ for the start-up of a
communication, and the time $\tau_c$ needed to send the data relative
to one particle to another processor.  The parameters represent the
fits to the data presented in $\S$\ref{sec:copy} and
$\S$\ref{sec:ringb}.

To measure the performance of our code and to compare the results with
the theoretical estimates, we consider the total {\it wall-clock time}
for an integration of one $N$-body time-unit
\cite{1986ussd.conf..233H}.  Our code employs ``standard N-body
units'' \cite{1986ussd.conf..233H}, according to which the
gravitational constant, the total mass, and the radius of the system
are taken to be unity. The resulting unit for time is called $N$-body
time-unit and is related to the physical crossing time of the system
through the relation $T_{\rm cross} = 2 \sqrt{2}$.  This time is
proportional to the number of particles in the system and represents
therefore a suitable unit to check the dependency of the execution
time on $N$.

\subsection{Performance of the copy algorithm}
\label{sec:copy}
The time needed for the computation of the force on a block of
particles of certain size can be estimated as follows.  Let $s$ be the
number of particles to be updated at a particular step (block size)
and let $s_i$ be the size of the subgroup of particles that processor
of rank $i$ has to update so that $s = \sum_{i=1}^{p} s_i$. The values
of $s_i$ will in general be different and the total time for the
computation will be determined by the processor with the largest block
size $s_{\rm max} = \max_{i=1,\dots p} \left\{s_i \right\}\th.$ If we
indicate with $\tau_f$ the time needed for one computation of the
force between two particles, then the time to compute the force on the
block of particles is given by $T_{\rm calc} = \tau_f N s_{\rm max}$,
since all the processors need to wait until the one with the largest
block of particles has terminated.  The time for the communication
taking place at the end of the computation is given by $T_{\rm comm} =
\tau_l p + \sum_{i=1}^{p} \tau_c s_i = \tau_l p + \tau_c s$, where
$\tau_l$ is the latency and $\tau_c$ is the time needed to send the
data relative to one particle from one processor to another.  The
total time to compute the force on the block of $s$ particles is
\begin{equation}
  \label{eq:timecopy}
  T_{\rm force} = T_{\rm calc} + T_{\rm comm} 
  = \tau_f N s_{\rm max} + \tau_l p + \tau_c s \th.
\end{equation}
Eq.\,(\ref{eq:timecopy}) shows how the time for the force
computation depends on the speed of calculation of each processor, on
the latency, and on the bandwidth of communication.  If we rewrite
$s_{\rm max} = s / p + \delta$, where the parameter $\delta$
represents the deviation from the mean value ($\delta \propto
\sqrt{s/p}$), then the calculation time scales as $T_{\rm calc}
\propto 1/p$ while the communication time scales as $T_{\rm comm}
\propto p$.

\begin{figure*}
 \begin{center}
   \includegraphics[width=6.5cm]{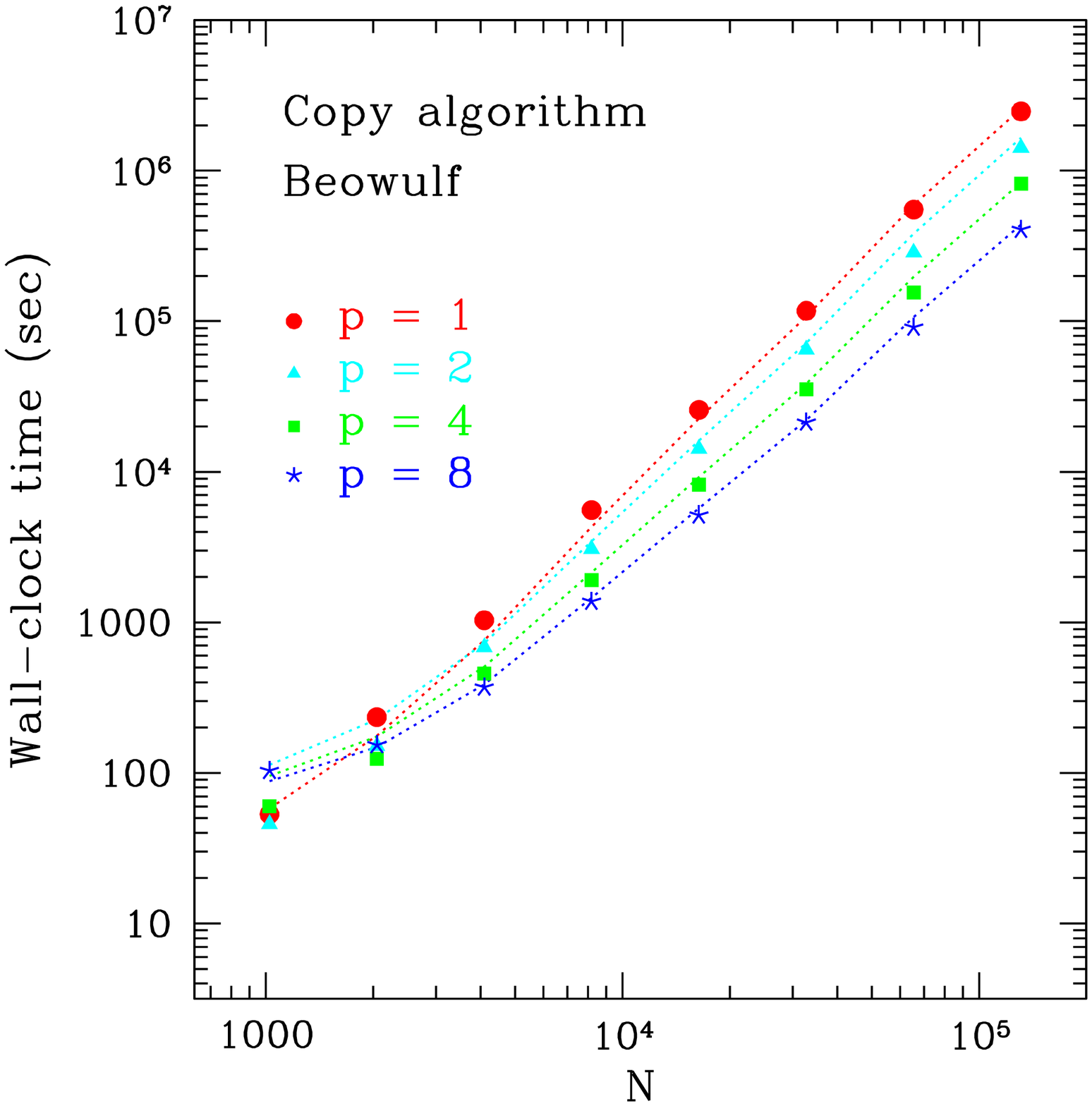}
   \includegraphics[width=6.5cm]{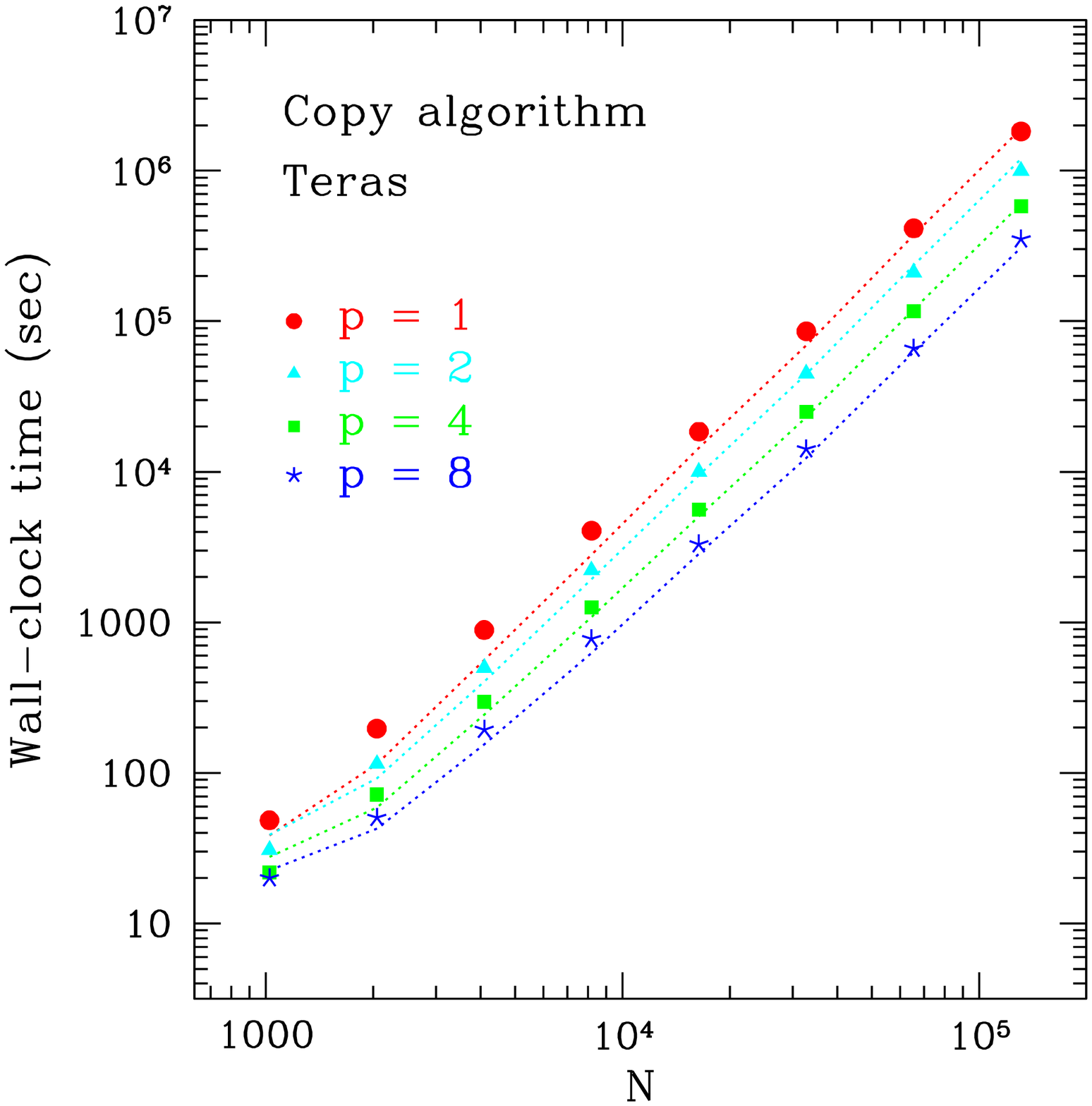}
   \includegraphics[width=6.5cm]{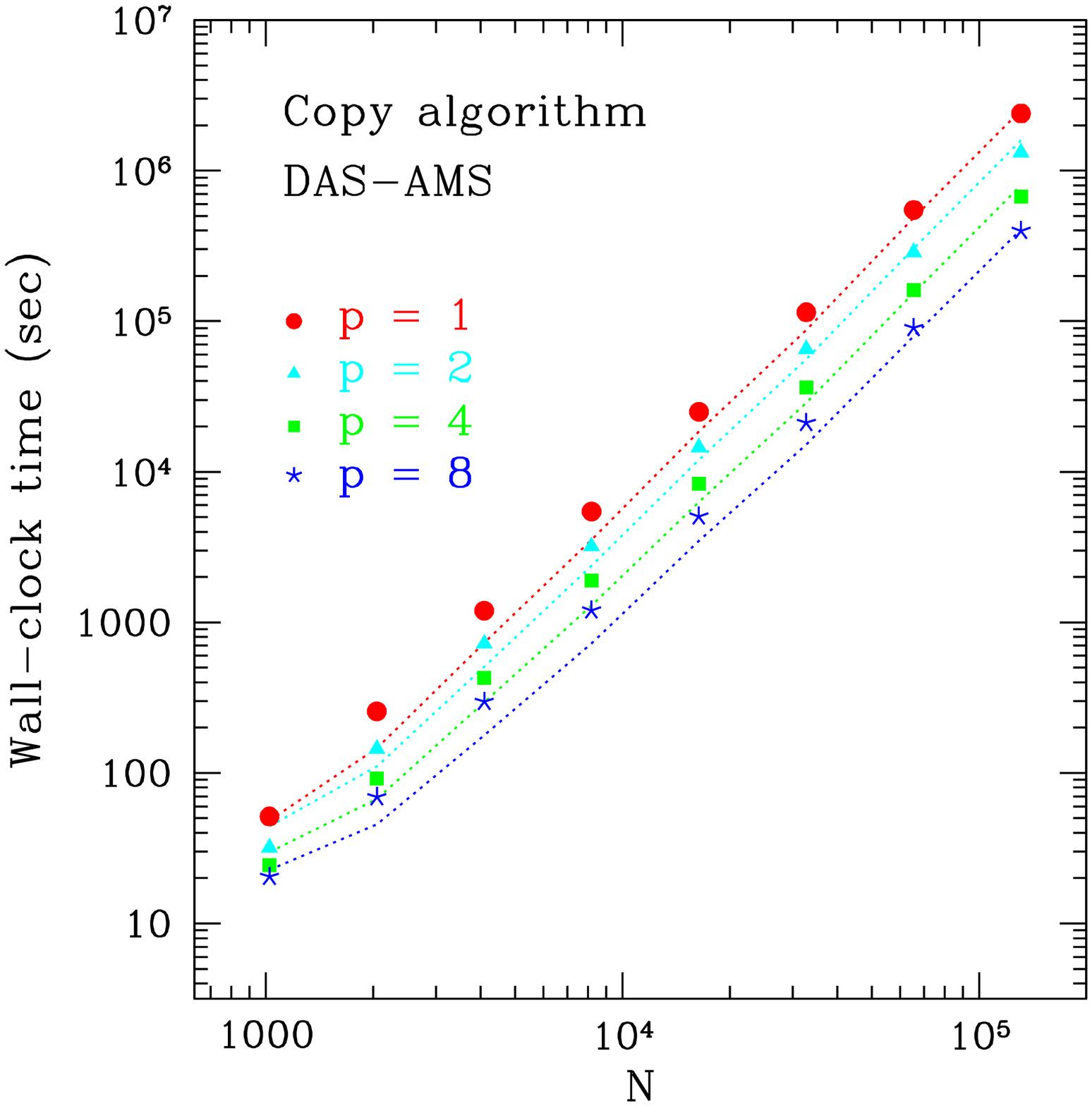}
   \includegraphics[width=6.5cm]{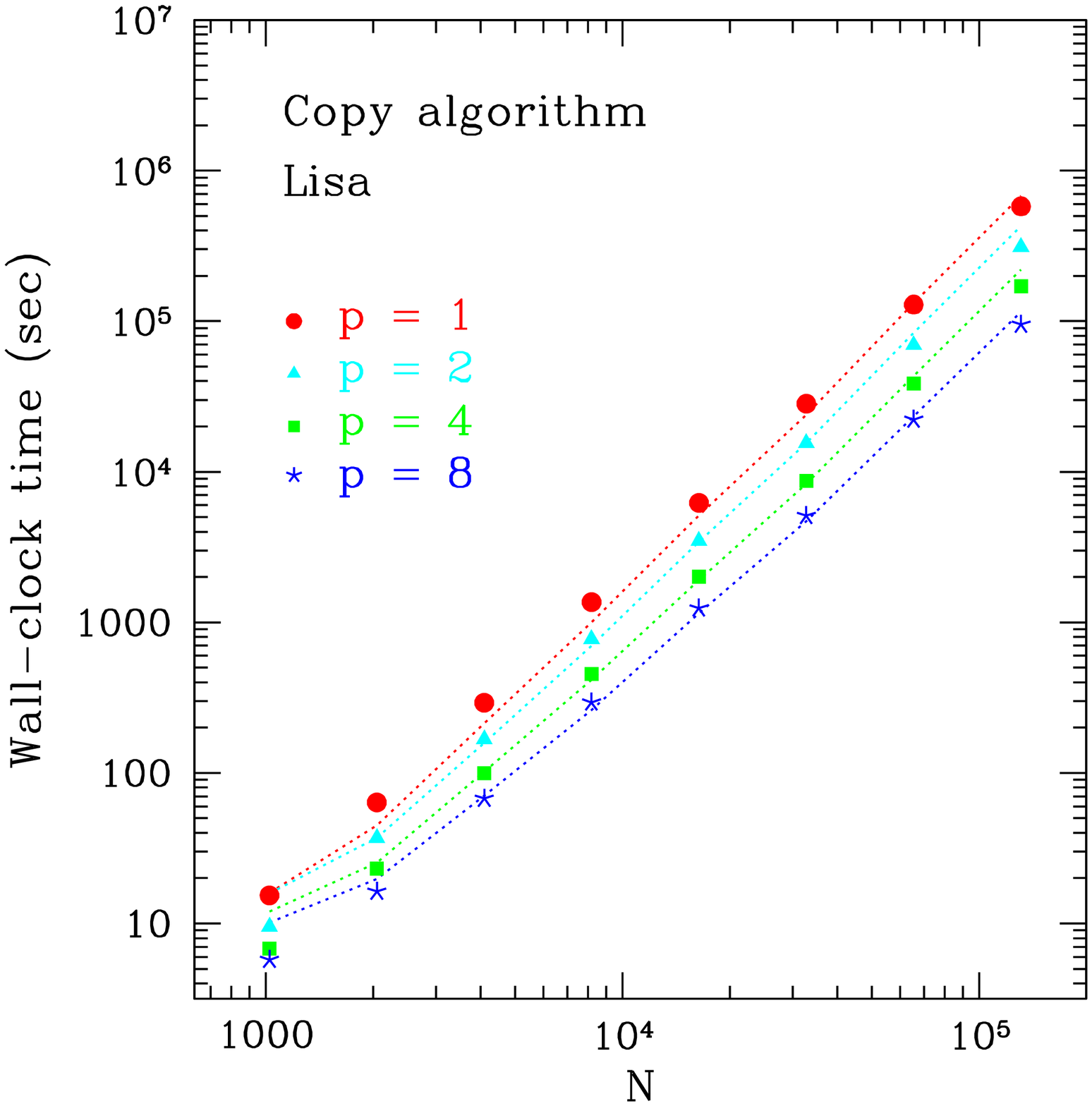}
   \caption{Wall-clock time as a function of the number of particles
     for the copy algorithm on the four different architectures.  The
     symbols represent the data obtained from the timing measurements
     while the dotted lines represent the predictions by the
     performance model.}
   \label{fig:plubkcp}
 \end{center} 
\end{figure*}

In Fig.\,\ref{fig:plubkcp} we report  the total wall-clock time for an
integration  of   a  Plummer  model\footnote{A  Plummer   model  is  a
spherically  symmetric  potential  used  to fit  observations  of  low
concentration  globular  clusters.} \cite{1911MNRAS..71..460P}  with
equal  mass stars  in virial  equilibrium for  one  $N$-body time-unit
using  the   copy  algorithm  on  the   four  different  architectures
(Table\,\ref{tab:computer})
The symbols represent the data obtained from the timing measurements
while the dotted lines represent the predictions by the theoretical
model.  Given the long execution times for large $N$, the measurements
shown in the plot were performed only once. Nonetheless, we performed
more measurements for a few cases to assess the typical uncertainties
associated with the timings and we found that the errors are always
smaller than 10\%. For large $N$, when the system is calculation
dominated, the performance is similar on the Beowulf, Teras and
DAS-AMS computers, while it is significantly better for the Lisa
cluster.  For small $N$, when the system is communication dominated,
the execution times are shorter on the clusters with faster network,
and this is especially true for a larger number of processors.  For a
fixed number of particles, the parallelization efficiency of the copy
algorithm reduces as the number of processors increases.  This is due
to the ever increasing amount of communication which takes place as
the number of processors becomes larger.  On the other hand, for a
fixed processor number, the efficiency increases as the number of
particles increases.  For larger $N$, in fact, the block sizes become
larger and the particles in the block tend to be more evenly
distributed among the available nodes (i.e. $\delta \rightarrow 0$).
Load imbalance can affect the performance of any parallel code and is
a result of the use of block time-steps.  Fortunately, for a large
number of particles a good load balance is achieved if the particles
are randomly assigned to the processors in the initialization phase.
The theoretical model used to predict the execution time is based on
Eq.\,(\ref{eq:timecopy}) with the parameters reported in
Table\,\ref{tab:computer}.  Since the block size is different at each
integration step, the total time for one $N$-body time-unit is
obtained considering the measured average block size $\langle s
\rangle \sim 0.1\, N^{2/3}$.  The agreement between the model and the
measured times is generally good and improves for large $N$, which is
the regime of interest for $N$-body simulations.

\subsection{Performance of the ring algorithm}
\label{sec:ringb}
In the case of the ring algorithm with blocking communication the
total time for one full force loop calculation can be estimated as
follows.  If we first consider the time for one shift in the ring, the
time to compute the force on the local subgroups of particles $s_i$ is
given by $T_{\rm calc, 1shift} = \tau_f n s_{\rm max}$, while the time
for communication is given by $T_{\rm comm, 1shift} = \tau_l + \tau_c
s_{\rm max}$.  After $p$ shifts in the ring, the total time to compute
the force on the block of $s$ particles is
\begin{equation}
  \label{eq:timeringb}
 T_{\rm force} = \left(T_{\rm calc, 1shift} + T_{\rm comm, 1shift}\right) p 
 = \tau_f N s_{\rm max} + \tau_l p + \tau_c p s_{\rm max}
\end{equation}
where we have substituted $n=N/p$.  As in the case of the copy
algorithm, the time for the force computation depends on the speed of
calculation of each processor, on the latency, and on the bandwidth of
communication.  A comparison of Eq.\,(\ref{eq:timecopy}) and
Eq.\,(\ref{eq:timeringb}) shows how the computation time required by
the two algorithms is exactly the same while the communication time is
generally slightly shorter for the copy algorithm. In the special case
$s_{\rm max} = s/p$, the communication time for the ring algorithm
equals that of the copy algorithm.

If the block size is the same for all the processors, for example $s_i
= s/p$, the total time for the force calculation is given by
\begin{equation}
  \label{eq:timebk}
  T_{\rm force} = \tau_f N \frac{s}{p} + \tau_l p + \tau_c s \th.
\end{equation}
The value $p_{\rm eq}$ of the number of processors for which the calculation
time and the communication time are equal is given by
\begin{equation}
\label{eq:peq}
p_{\rm eq} = \frac{- \tau_c s + \sqrt{\tau_c^2 s^2 + 4 \tau_l \tau_f N s}}
{2 \tau_l} \th.
\end{equation}
Since the calculation time monotonically decreases as a function of
$p$ while the communication time monotonically increases as a function
of $p$, there exists a specific value $p_{\rm min}$ for the number of
processors which minimizes the total time. Solving for the minimum
yields $p_{\rm min} = \sqrt{\frac{\tau_f}{\tau_l} N s} \th$.  For
moderately concentrated models, $s \propto N^{2/3}$
\cite{1991PASJ...43..859M} and hence $p_{\rm min}\propto N^{5/6}$.  In
the simple case $s_i = s/p$, the time to compute the force exerted on
the block of particles is the same for the copy and the ring
algorithms, and therefore the expressions for $p_{\rm eq}$ and $p_{\rm
min}$ hold for both schemes. In the more general case of different
$s_i$, the expressions for $p_{\rm eq}$ and $p_{\rm min}$ differ
slightly.

\begin{figure}
  \begin{center}
    \includegraphics[width=6cm]{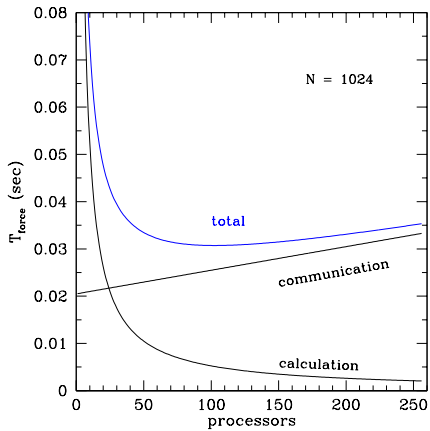}
    \caption{Theoretical estimate of the total time for one full force
      loop obtained from Eq.\,(\ref{eq:timebk}) in the simple case
      $N=1024$, $s_i= N/p$ for each processor, using the parameters
      for a Beowulf cluster (see Table\,\ref{tab:computer}).  The
      calculation time scales as $T_{\rm calc} \propto 1/p$ while the
      communication time scales as $T_{\rm comm} \propto p$ so that
      there exists a particular value of the processor number for
      which the two times are equal. In this case $p_{\rm eq} = 24$.
      The number of processors in correspondence of which the total
      time is minimum is $p_{\rm min} = 102$.}
    \label{fig:timebk}
  \end{center}
\end{figure}

In the case of a shared time-step code the block size is the same for
all the processors and is given by $s_i = n$. The total time for the
force calculation becomes
\begin{equation}
  \label{eq:timesh}
  T_{\rm force} = \tau_f \frac{N^2}{p} + \tau_l p + \tau_c N
\end{equation}
so that for a fixed number of particles $T_{\rm calc} \propto 1/p$
and $T_{\rm comm} \propto p$.
Fig.\,\ref{fig:timebk} shows the calculation and communication time 
as a function of the number of processors for a fixed number of particles.
The total time has a minimum in correspondence of $p_{\rm min}=
\sqrt{\tau_f/\tau_l} N$ while the calculation and communication time
are equal for $p_{\rm eq} = \frac{N}{2\tau_l} \left(-\tau_c +
\sqrt{\tau_c^2 + 4 \tau_f \tau_l}\right)$.  A shared time-step code
allows the use of a larger number of processors compared to a block
time-step code for the same efficiency because of the larger block
size.

To validate the model we compare the execution time predicted by
Eq.\,(\ref{eq:timesh}) with the results obtained integrating a shared
time-step code for one step.  The prediction is accurate to a level of
10-20\% for the range $N$ = 1024 - 16384.  The theoretical prediction
of the execution time for a block time-step code is complicated by the
fact that the block size changes with time.  Eq.\,(\ref{eq:timebk})
can be satisfactorily applied to predict the time $T_{\rm force}$ at a
specific step only if the value of the block size is known.
Nonetheless, by assuming an average block size $\langle s \rangle \sim
0.1\, N^{2/3}$, we can apply the performance model to the block
time-step code and predict the execution over one $N$-body time-unit
fairly accurately.

\begin{figure*}
  \begin{center}
    \includegraphics[width=6.5cm]{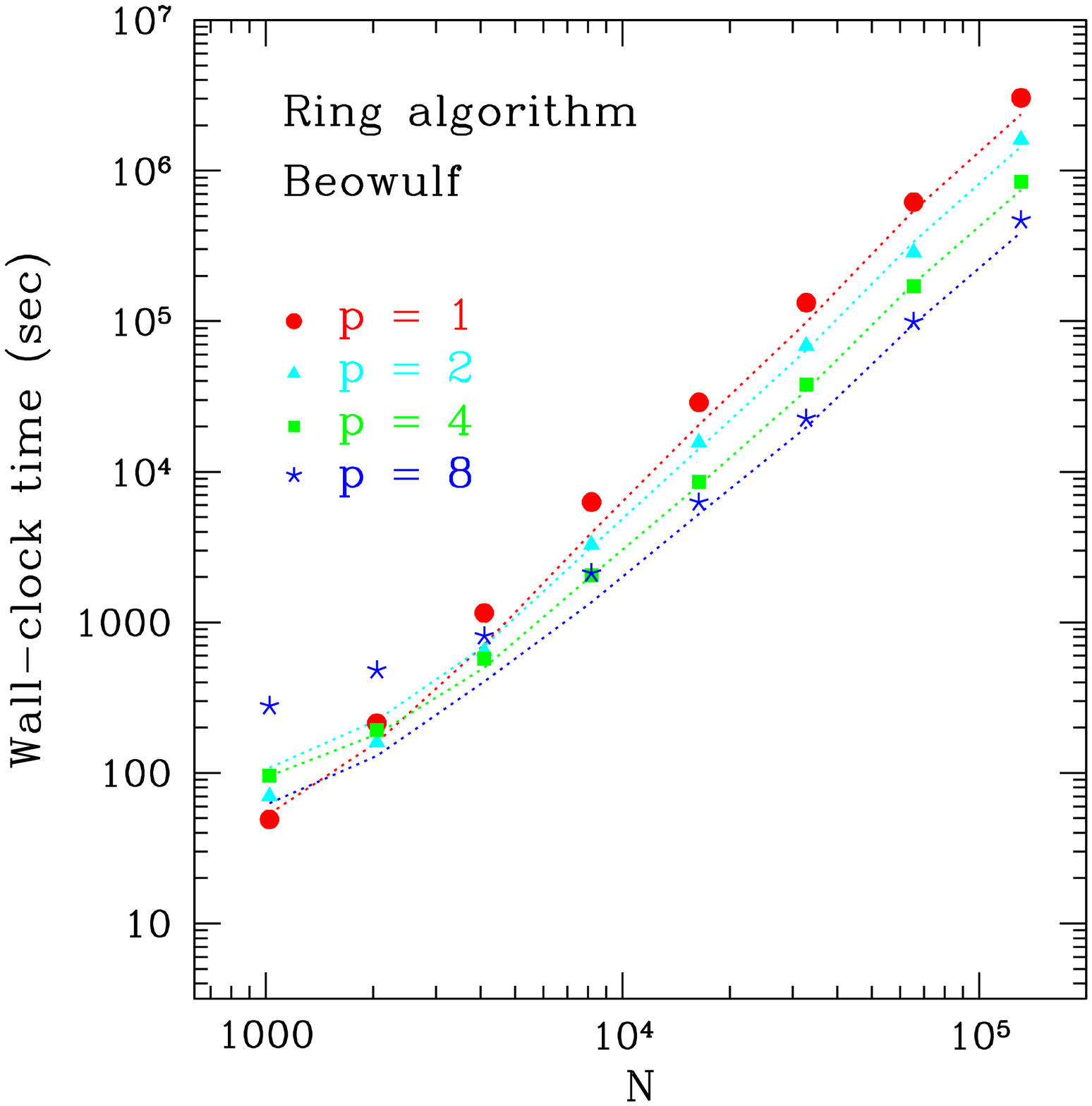}
    \includegraphics[width=6.5cm]{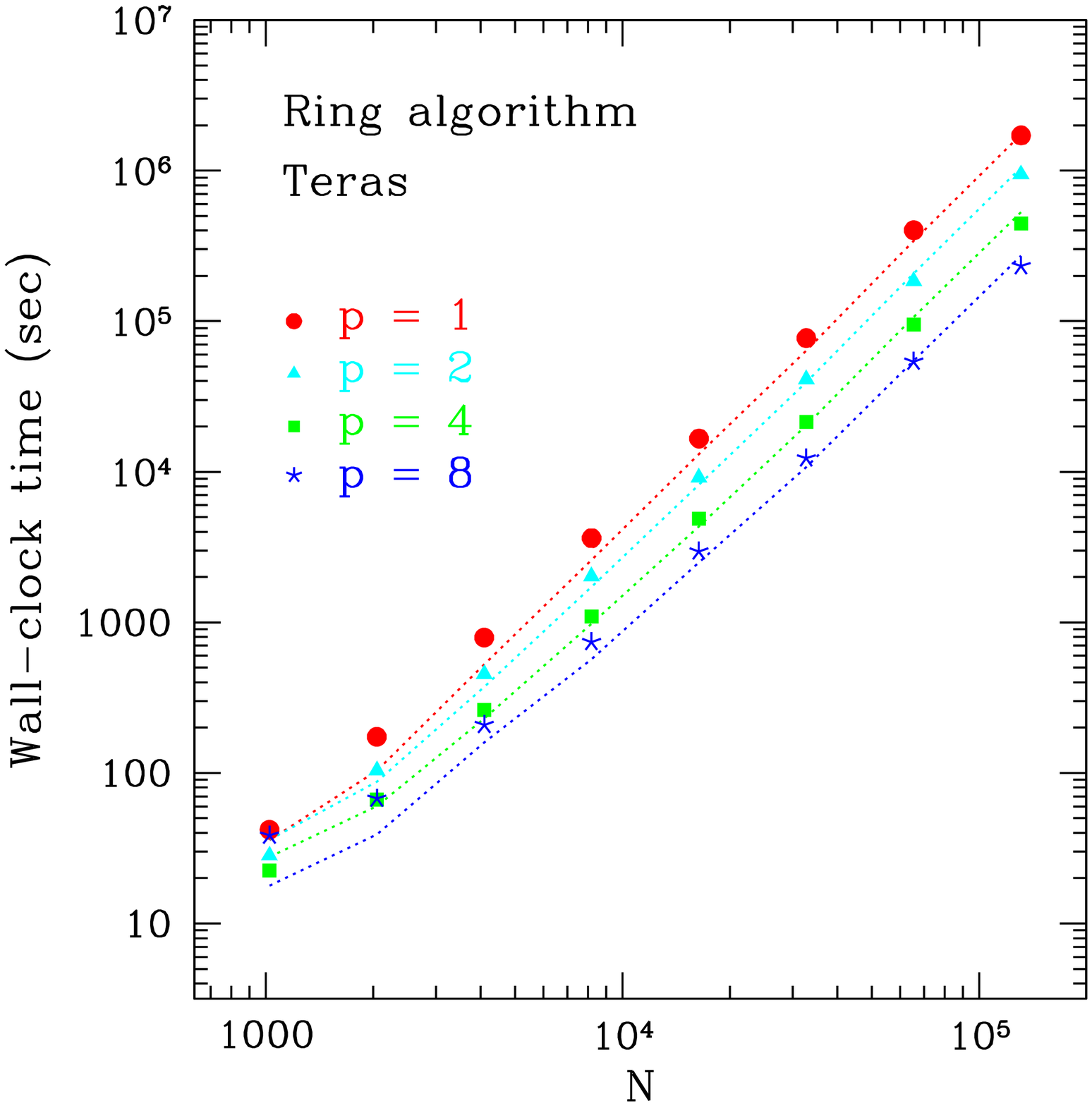}
    \includegraphics[width=6.5cm]{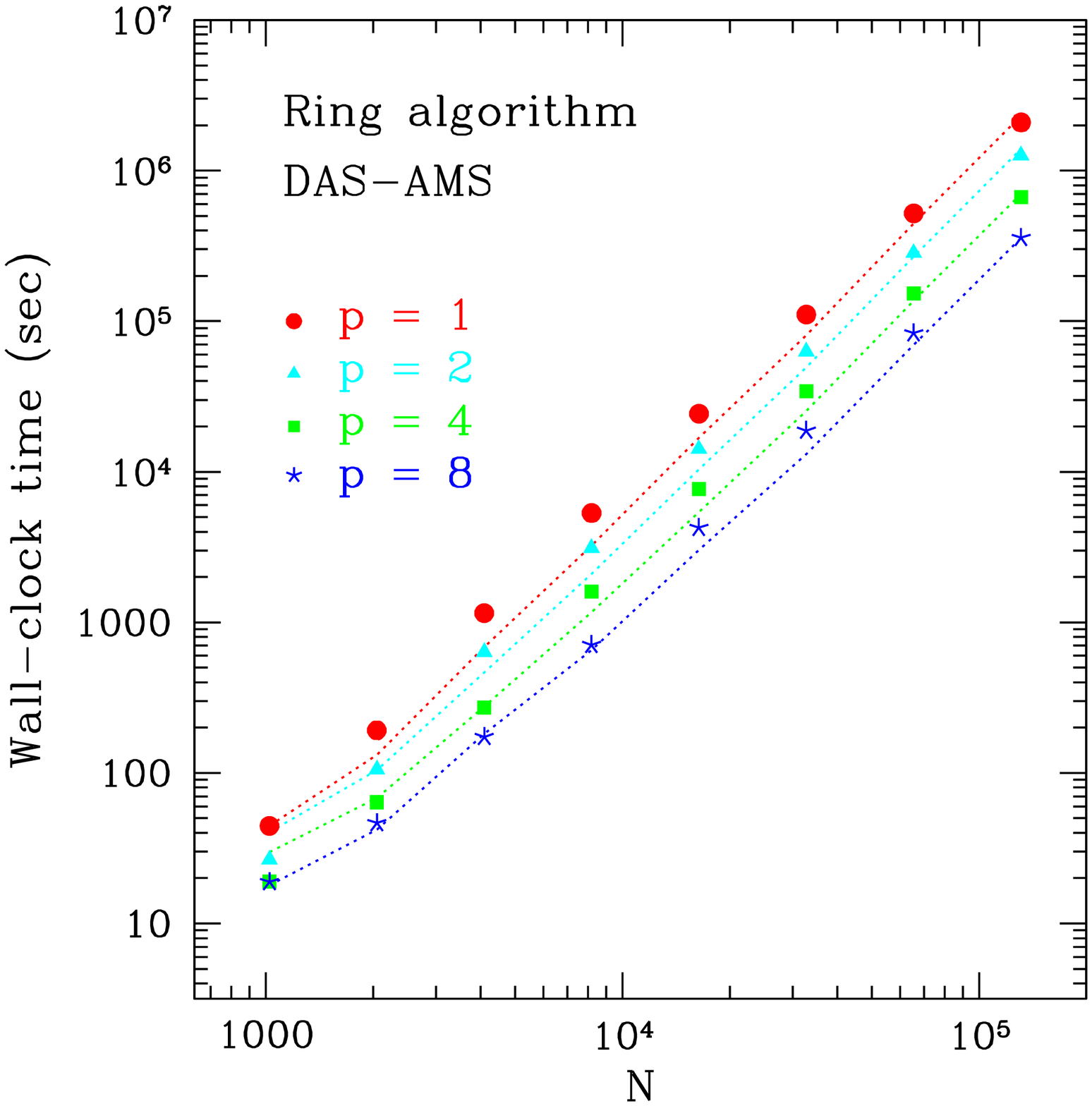}
    \includegraphics[width=6.5cm]{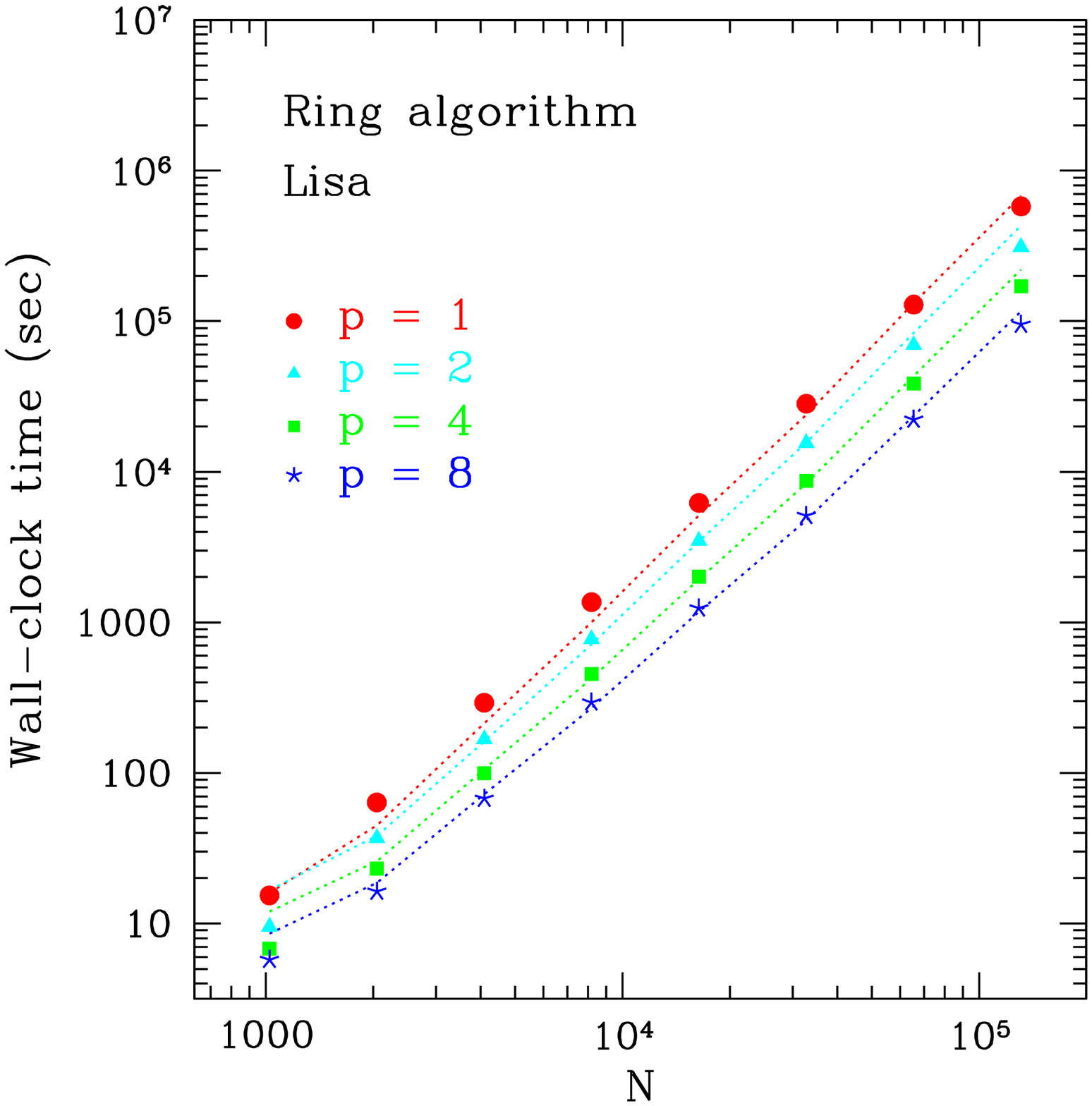}
    \caption{Wall-clock time as a function of the number of particles
      for the ring algorithm with blocking communication on the
      different architectures. The symbols represent the data obtained
      from the timing measurements while the dotted lines represent
      the predictions by the performance model.}
    \label{fig:plubkrng}
  \end{center}
\end{figure*}

In Fig.\,\ref{fig:plubkrng} we report the total wall-clock time for
the integration of the block time-step code using the ring algorithm
on the four different architectures (Table\,\ref{tab:computer}).
The symbols represent the data obtained from the timing measurements
while the dotted lines represent the predictions by the theoretical
model.  The particles are initially distributed in space according to
a Plummer model with equal mass stars in virial equilibrium and the
integration is for one $N$-body time-unit.  The ring and the copy
algorithm have a similar performance in terms of total execution time
for large numbers of particles, whereas the ring algorithm is heavily
dominated by communication for small numbers of particles.  Like in
the case of the copy algorithm, the efficiency decreases for large
numbers of processors, where the communication governs the general
performance, but increases for large numbers of particles.  The
agreement between the model and the measured times is generally good,
especially for large $N$.

\begin{figure}
  \begin{center}
    \includegraphics[width = 7cm]{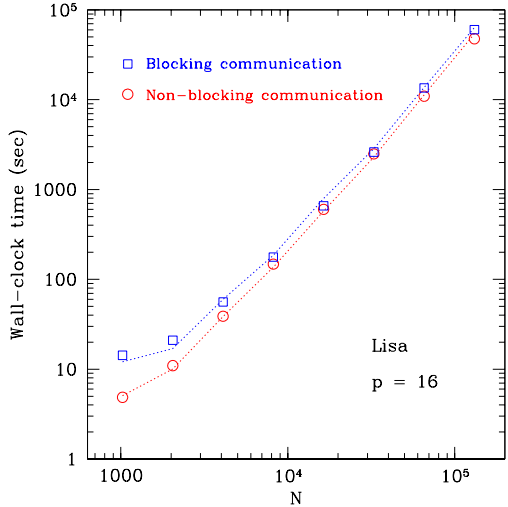}
    \caption{Timing comparison between the blocking and non-blocking
      communication ring algorithm.  The runs were performed on the
      Lisa cluster with $p = 16$ using a Plummer model and the
      integration time was one $N$-body time-unit in all cases.  The
      symbols represent the data obtained from the timing measurements
      while the dotted lines represent the predictions by the
      performance model.  For small $N$ the non-blocking scheme
      reduces the execution time by about a factor two, presenting an
      almost linear scaling.  For larger $N$, where the total time is
      dominated by the force computation, both schemes present an
      $O(N^2)$ scaling with the number of particles and achieve a
      similar performance.}
    \label{fig:plbnb}
  \end{center} 
\end{figure}

If non-blocking communication is used for the ring algorithm
\cite{775815}, the total execution time for one full force calculation
can be significantly reduced.  The calculation time for one shift of
the systolic loop is the same as in the blocking case: $T_{\rm calc, 1
shift} = \tau_f n s_{\rm max}\th$.  The communication time has two
separate contributions, one from the transfer of the positions and
velocities and one from the transfer of the accelerations and jerks.
We define $\tau_{pv}$ as the time needed to send the position and the
velocity vectors of one particle to another processor.  Similarly, we
define $\tau_{aj}$ as the time needed to send the acceleration and the
jerk vectors of one particle to another processor.  Since the two
communications are taking place simultaneously, the total
communication time is given by the maximum between the two: $T_{\rm
comm, 1 shift} = {\rm max} \left[ T_{\rm comm,1} , T_{\rm comm,2}
\right]$, where $T_{\rm comm,1} = \tau_l + \tau_{pv} s_{\rm max}$ is
the time needed to transfer the positions and velocities of the block
of particles while $T_{\rm comm,2} = \tau_l + \tau_{aj} s_{\rm max}$
is the time needed to transfer the accelerations and jerks of the same
block.  At the end of the last shift an additional communication is
required $T_{\rm comm,final} = T_{\rm comm,2}\th$.  After $p$ shifts
in the ring the total time to compute the force on the block of $s$
particles is
\begin{equation}
  \label{eq:timeringnb}
  T_{\rm force} = {\rm max} \left[T_{\rm calc, 1 shift}, 
T_{\rm comm, 1 shift}\right] p   + T_{\rm comm,final}\th.
\end{equation}
The use of non-blocking communication is efficient whenever the
communication time is not negligible compared to the calculation
time. For moderately concentrated models like the Plummer model this
happens for small numbers of particles, when the average block size is
small and hence the particles are less likely to be evenly distributed
(see Fig.\,\ref{fig:plbnb}).

\section{Performance on the BlueGene/L supercomputer}
The BlueGene/L supercomputer is a novel machine developed by IBM to
provide a very high number of computing nodes with a modest power
requirement.  Each node consists of two processors, a special variant
of IBM's Power family, with a clock speed of 700 Mhz.  To obtain good
performance at this relatively low frequency, each node processes
multiple instructions per clock cycle.  The nodes are interconnected
through multiple complementary high-speed low-latency networks,
including a 3D torus network and a combining tree network.

\begin{figure}
  \begin{center}
    \includegraphics[width = 7cm]{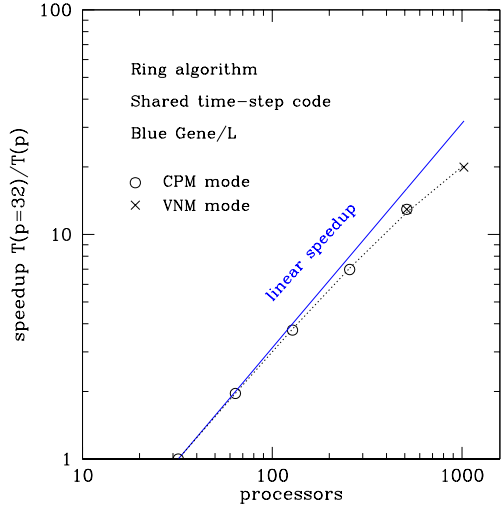}
    \caption{BlueGene/L speedup over 32 nodes in the case of the
      integration of a $N$=131072 Plummer model with a shared
      time-step code for one step.  The full dots are the results for
      the Coprocessor mode (CPM) while the full square is the result
      for the Virtual Node Mode (VNM) (see \cite{almasi2004} for
      technical details).  The dotted line represents the prediction
      by Eq.\,(\ref{eq:timesh}).  }
    \label{fig:bg}
  \end{center}
\end{figure}

We could perform a limited number of test runs on the Blue Gene/L
machine hosted by the IBM Watson research center.  We evolved an
$N$=32768 Plummer model for a time $t$ = 0.03125 time-units using the
block time-step code parallelized with the non-blocking ring
algorithm.  The execution times on 32, 64, 128 processors were 320
sec, 256 sec, 222 sec, respectively. Consequently, the speedup on 64
and 128 processors, relative to 32 nodes, was of only 1.25 and 1.5
respectively.  We realized that a block time-step code is not
efficient for a combination of a relatively small number of particles
and a large number of processors.  We then evolved a $N$=131072
Plummer model for one time-step using a shared time-step code
parallelized with the ring algorithm.  In Fig.\,\ref{fig:bg} we show
the timing results, together with the predictions by the theoretical
model, as a function of the number of processors.  An almost linear
speedup is achieved by means of an efficient use of both processors in
a node \cite{almasi2004}, with a peak speed of 2.8 GFlop/s per node.
The theoretical model for a shared time-step code, which is based on
Eq.\,(\ref{eq:timesh}), provides a very accurate prediction of the
execution time for the following set of parameters: $\tau_f =
1.5\times10^{-7}\,s$, $\tau_l = 5.0\times10^{-6}\,s$, $\tau_c =
1.0\times10^{-5}\,s$.
These short test runs show that the Blue Gene/L supercomputer can
achieve good performance and almost optimal speedup under conditions
of good load balance.

\section{Performance analysis on the GRID}
Grid technology is rapidly becoming a major component of computational
science.  It offers a unified means of access to different and distant
computational resources, with the possibility to securely access
highly distributed resources that scientists do not necessarily own or
have an account on.  Connectivity between distant locations,
interoperability between different kinds of systems, and resources and
high levels of computational performance are some of the most
promising characteristics of the Grid.  Although significant
improvement in the performance of direct codes can be obtained by
means of general purpose parallel computers (see $\S$
\ref{sec:parallel}), the use of highly distributed clusters within
computational grids has not yet been explored.

In this section we present the results of experiments conducted on
computational grids using the $N$-body code parallelized with a
systolic algorithm and the MPICH-G2 device across large geographical
distances.  We explore the total effects of network latency on the
performance on the 5-cluster DAS-2\footnote{\tt http://www.cs.vu.nl/das2} Grid, distributed within the Netherlands and running the Globus toolkit \cite{fk97}, as well as on the 18-node CrossGrid\footnote{\tt http://www.eu-Crossgrid.org} testbed, distributed across Europe and running LCG2\footnote{\tt http://lcg.web.cern.ch/LCG/Documents/default.htm}.

The MPI implementation that is used in a Globus environment is
MPICH-G2, which is a grid-enabled implementation of the standard MPI
v1.1.  Using services from the Globus Toolkit, MPICH-G2 allows one to
couple multiple machines, potentially of different architectures, to
run MPI applications.  It automatically converts data in messages sent
between machines of different architectures and supports
multi-protocol communication by automatically selecting TCP for
inter-machine messaging and vendor-supplied MPI for intra-machine
messaging.

We now describe the two grid testbeds used for our experiments.

\subsection{The DAS-2 and CrossGrid testbeds}
The Distributed ASCI Supercomputer (DAS-2) is a wide-area parallel
computer which consists of clusters of workstations distributed across
the Netherlands.  The DAS-2 virtual machine is used for research on
parallel and distributed computing by five Dutch universities and
contains 200 computing nodes in total.  Each node contains two 1 GHz
Pentium IIIs, at least 1 GB RAM and a 20 GB local IDE disk.  The nodes
within a local cluster are connected by a Myrinet-2000 network, which
is used as high-speed interconnection, while Fast Ethernet is used as
OS network.  The five local clusters are connected by Surfnet, the
Internet backbone for wide-area communication across universities in
the Netherlands.  The version of MPICH-G2 available on DAS-2 is
MPICH-GM, which uses Myricom's GM as its message passing layer on
Myrinet.

\subsection{The CrossGrid testbed}
The CrossGrid pan-European distributed testbed shares resources across
16 European sites.  The sites range from relatively small computing
facilities in universities to large research computing centers,
offering an ideal mixture to test the possibilities of an experimental
Grid framework.  National research networks and the high-performance
European network, Geant \cite{geant}, assure inter-connectivity
between all sites.  The network includes a local step via Fast or
Gigabit Ethernet, a jump via a national network provider at speeds
that will range from 34 Mbits/s to 622 Mbits/s or even Gigabit and a
link to the Geant European network at 155 Mbits/s to 2.5 Gbits/s.  The
platforms include Intel Pentium III and Intel Xeon processors with
speeds ranging from 1 to 2.4 GHz.

The CrossGrid team focuses on the development of Grid middle-ware
components, tools, and applications with a special focus on parallel
and interactive computing, deployed across 11 countries. The added
value of this project consists in the extension of the Grid to support
interactive applications.  The CrossGrid testbed largely benefits from
the European Data Grid \cite{fkt01} experience on testbed setup and
Globus middle-ware distributions.

\subsection{Performance results}
As shown in $\S$ \ref{sec:parallel}, the main factors determining the
general performance of a parallel application are the calculation
speed of each node, the bandwidth of the inter-processor communication
and the network latency.  In the case of a computational grid, the
latency between different clusters and the slower network may sensibly
affect the execution times.

Adopting the same nomenclature as in $\S$ \ref{sec:copy} and
\ref{sec:ringb}, we derive a theoretical estimate for the time $T_{\rm
force}$ in the most general case of a heterogeneous grid, where each
processor has a different CPU speed and any pair of processors is
interconnected by a different network.  In the case of the ring
algorithm with blocking communication the total time after one shift
is given by $T_{\rm force, 1 shift} = \max_{i=1 \dots p}
\left\{\tau_{f, i}~n~s_i\right\} + \max_{i=1 \dots p} \left\{\tau_{l,
i} + \tau_{c,i}~s_i\right\}$, where the subscript $i$ refers to
processor $i$.  Taking into account the fact that each processor may
have a different block size $s_i$, the total time needed for the
computation of the forces exerted on the block of $s$ particles can be
written as
\begin{equation}
\label{eq:timegrid}
T_{\rm force} = \sum_{shift=1}^{p} n ~ \max_{i=1 \dots p} 
\left\{\tau_{f, i}~s_i\right\} + \sum_{shift=1}^{p} \max_{i=1 \dots p} 
\left\{\tau_{l, i} + \tau_{c,i}~s_i\right\}~.
\end{equation}
In the ideal case of all processors having the same block size $s_i = s/p$,
the previous equation simplifies to
\begin{equation}
 T_{\rm force} = N ~ \frac{s}{p} \max_{i=1 \dots p} \left\{\tau_{f, i}\right\} + 
 p~\max_{i=1 \dots p} \left\{\tau_{l, i} + \tau_{c,i}~\frac{s}{p}\right\}~.
\end{equation}

To measure the effect of latency we performed several test runs on the
DAS-2 low latency supercomputer and the CrossGrid testbed using the
systolic $N$-body code.  The specifications for the DAS-2
\cite{Iskra2003} and for the CrossGrid are shown in Table\,\ref{tab:gridspec}.
\begin{table}[h]
  \caption{Specifications for the grid testbeds.}
  \label{tab:gridspec}
  \centering
    \begin{tabular}{cccccc}
      \hline
      & CPU  & Network & $\tau_f$ [sec] & $\tau_l$ [sec] & $\tau_c$ [sec]  \\
      \hline
      \hline
      DAS-2      & 1 GHz & Surfnet  & 4.5$\times10^{-7}$ & 5.0$\times10^{-5}$ & 8.0$\times 10^{-4}$ \\
      CrossGrid  & 1 GHz & Geant    & 4.5$\times10^{-7}$ & 2.0$\times10^{-3}$ & 4.0$\times 10^{-3}$ \\
      \hline
    \end{tabular}
\end{table}

We evolved the same initial configuration (Plummer model) for one
$N$-body time-unit using 4 processors.
\begin{figure}
  \begin{center}
    \includegraphics[width = 7cm]{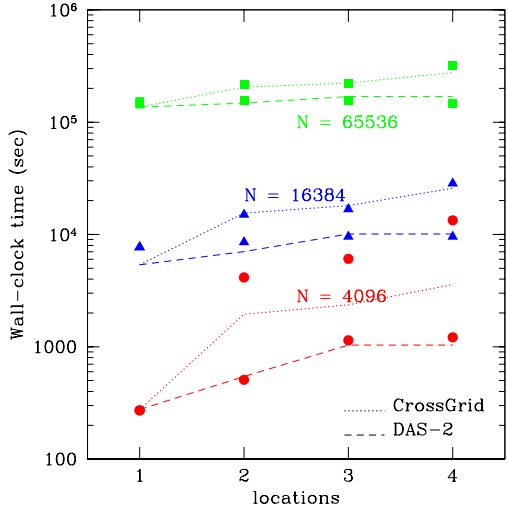}
    \caption{Performance comparison of the direct $N$-body code 
      parallelized with the ring algorithm on the DAS-2 wide-area supercomputer 
      (dashed lines) and the CrossGrid distributed testbed (dotted lines).
      The wall-clock time is plotted as a function of the number of locations
      where the computing nodes are selected.
      The timing refers to the integration of a Plummer model for N=4096
      (full dots), 16384 (full triangles), 65536 (full squares), 
      for one $N$-body time-unit. The dotted and dashed lines indicate 
      the predictions by the performance model Eq.\,(\ref{eq:timegrid}) 
      .}
    \label{fig:perfgrid}
  \end{center}
\end{figure}
The total execution time is plotted in Fig.\,\ref{fig:perfgrid} as a
function of the number of different locations hosting the computing
nodes.  The low latency network on the DAS-2 generally results in a
good performance even if the nodes are allocated in different
clusters.  Only in the case of a very small number of particles, like
for the $N$ = 4096 run, the execution time increases steadily with the
number of locations.  This is due to an unfavorable computation to
communication ratio for small $N$. The effects of inter-process
communication are more evident for the CrossGrid runs, where the
execution time increases substantially with the number of locations.
However, the performance improves as the size of the $N$-body system
increases since the computation to communication ratio becomes higher
and a better load balance can be achieved.  For large systems, the
performance on the CrossGrid is at most a factor three worse than that
on DAS-2.  The performance model can predict the execution time for
DAS-2, which is an homogeneous system, while can only reproduce the
behavior of the CrossGrid for large $N$, when the calculation
dominates over communication.  Even though the clusters have been
selected to be as similar as possible, the CrossGrid is not an
homogeneous system and the different distances between the locations
result in different communication times.

\section{Discussion}
A numerical challenge for the astronomical community in the next years
will be the simulation of star clusters containing one million stars.
We have shown that direct $N$-body codes can efficiently be applied to
the simulation of large stellar systems and that their performance can
be predicted with simple models.

In this section, we apply the performance model introduced in
\S\,\ref{sec:copy} for the copy algorithm to predict the total
execution time for the simulation of a system with $N=10^6$. 
In Table\,\ref{tab:pred} we report our predictions for the execution
times in the simulation of such a star cluster.  For the prediction we
consider the parameters of current state-of-the-art supercomputers
(Lisa and BlueGene) and of an expected state-of-the-art supercomputer
in ten years time (Future).  For the Future computer we consider the
Lisa cluster as a reference and we assume that the CPU speed doubles
every 18 months the network speed doubles every 9 months.  The
resulting parameters for the Future computer are $\tau_f =
2.0\times10^{-9}$, $\tau_c = 1.0\times10^{-6}$ and $\tau_l =
5.0\times10^{-7}$.
\begin{table}[h]
  \caption{Predicted execution time for a star cluster.}
  \label{tab:pred}
  \centering
    \begin{tabular}{ccrcc}
      \hline
      System   & N & $p_{\rm min}$ & time (1 $N$-body unit) & time (1000 $N$-body units)\\
      \hline
      \hline
      Lisa     & $10^6$ &   1000 &  1.3 days & 3.5 years\\
      BlueGene & $10^6$ &   1000 &  1.0 day  & 3.1 years\\
      Future   & $10^6$ &   1000 &  30 min   & 12 days\\
      \hline
    \end{tabular}
\end{table}

We find that a typical supercomputer in ten years will be able to
simulate a star cluster (one million stars) for one $N$-body time-unit
in about thirty minutes using 1000 processors.  A full simulation of
1000 $N$-body units will thus require less than a month to complete.
Large supercomputers containing at least 1000 processors will
be necessary to perform the first realistic simulation of a large globular clusters.
Algorithmic developments are unlikely to result in a reduction
of the calculation time as the $N^2$ operation will always remain
the most demanding part of a simulation using a direct code.
On the other hand, new treatments of the additional physics
involved in a realistic simulation (stellar and binary evolution,
dynamical encounters, collisions between stars) will be very 
important for future comparisons between observations and simulations
of star clusters.

The simulation of larger systems like galactic nuclei ($10^9$ stars)
or galaxies ($10^{11}$ stars) will be unrealistic to perform on a large
supercomputer, even with future special purpose hardware.  
The best method for this objective is the use of hybrid codes 
in which a direct integration is combined with less accurate 
but faster techniques, like Monte Carlo codes or tree codes.
The different dynamical evolution of large systems like galaxies, 
which can be considered ``non-collisional'', will allow for the 
partial use of approximated methods without loss of fundamental
physical phenomena.

\section{Conclusions}
We have implemented two parallelization schemes for direct $N$-body
codes with block time-steps, the copy and ring algorithm, and compared
their performance on different parallel computers.  In the case of
clusters or supercomputers, the execution times for the two schemes
are comparable except for very small systems where the communication
time dominates over the calculation time and hence the copy algorithm
performs slightly better.  The ring algorithm is well suited for the
integration of very large systems, because of its reduced memory
demands, and for very concentrated models, where the average block
size becomes very small and the algorithm can be implemented with
non-blocking communication to limit the effects of load imbalance.

The timing experiments we have conducted on two Grid testbeds indicate
that the performance on large grids is not significantly worsened by
the communication among nodes residing in different locations,
provided that the size of the $N$-body system is sufficiently large to
ensure a high computation to communication ratio and a good load
balance.  Although these results are only preliminary, they appear
very promising in the direction of ever larger $N$-body simulations of
star clusters.

We have developed a performance model for each parallel scheme and we
have applied it to the prediction of the execution time for the
simulation of a star cluster containing one million stars.  We expect
such simulation to become feasible on a supercomputer in ten
years. Simulating entire galaxies however is not foreseeable in the
near future without major software developments.

\section {Acknowledgments}
We thank Jun Makino and Piet Hut for support through the ACS project
({\tt http://www.artcompsci.org}).  We also thank Gyan Bhanot, Bob
Walkup and the IBM T.J Watson Research Center for performing test runs
on the BlueGene supercomputer, the DAS-2 and CrossGrid projects, for
their technical help in the grid tests and Rainer Spurzem and David
Merritt for useful discussions on parallel computing.  This work was
supported by the Netherlands Organization for Scientific Research (NWO
grant \#635.000.001), the Royal Netherlands Academy of Arts and
Sciences (KNAW) and the Netherlands Research School for Astronomy
(NOVA).

\bibliographystyle{elsart-num}
\bibliography{nbody.bib}

\end{document}